\documentclass[12pt]{article}
\usepackage{amsmath}
\usepackage{amssymb}

\numberwithin{equation}{section}
\newcommand{\kb}{\bar{K}}

\newcommand{\veff}{V_{\rm{eff}}}
\newcommand{\vteff}{\tilde{V}_{\rm{eff}}}
\newcommand{\psia}{\psi^{(\alpha)}}
\newcommand{\ea}{E^{(\alpha)}}
\newcommand{\na}{{\cal N}^{(\alpha)}}
\newcommand{\lambdaa}{\frac{\lambda}{\alpha}}
\newcommand{\kta}{\tilde{K}^{(\alpha)}}
\newcommand{\qa}{Q^{(\alpha)}}
\newcommand{\aal}{a^{(\alpha)}}
\newcommand{\ba}{b^{(\alpha)}}
\newcommand{\ga}{g^{(\alpha)}}
\newcommand{\kba}{\bar{K}^{(\alpha)}}
\newcommand{\ka}{K^{(\alpha)}}
\newcommand{\N}{\mathbb{N}}
\newcommand{\Aa}{A^{(\alpha)}}
\newcommand{\ca}{C^{(\alpha)}}

\title{
Spectrum generating algebras for position-dependent mass oscillator Schr\"odinger equations}
\author{C Quesne\\ 
{\small Physique Nucl\'eaire Th\'eorique et Physique Math\'ematique,  Universit\'e Libre de Bruxelles,} \\ 
{\small Campus de la Plaine CP229, Boulevard~du Triomphe, B-1050 Brussels, Belgium}}
\date{ }
\begin{document}
\baselineskip=22pt plus 1pt minus 1pt
%%%%%%%%%%%%%%%%%%%%%%%%%%%%%%%%%%%%%%%%%%%%%%%%%%%%%%%%%%
\maketitle

\begin{abstract} 
The interest of quadratic algebras for position-dependent mass Schr\"odinger equations is highlighted by constructing spectrum generating algebras for a class of $d$-dimensional radial harmonic oscillators with $d \ge 2$ and a specific mass choice depending on some positive parameter $\alpha$. Via some minor changes, the one-dimensional oscillator on the line with the same kind of mass is included in this class. The existence of a single unitary irreducible representation belonging to the positive-discrete series type for $d \ge 2$ and of two of them for $d=1$ is proved. The transition to the constant-mass limit $\alpha \to 0$ is studied and deformed su(1,1) generators are constructed. These operators are finally used to generate all the bound-state wavefunctions by an algebraic procedure.
\end{abstract}

\noindent
Short title: Spectrum generating algebras


%
%========================================================================
%
\newpage
\section{Introduction}

During recent years, quantum mechanical systems with a position-dependent (effective) mass (PDM) have attracted a lot of attention and inspired intense research activites. They are indeed very useful in the study of many physical systems, such as electronic properties of semiconductors~\cite{bastard} and quantum dots~\cite{serra}, nuclei~\cite{ring}, quantum liquids~\cite{arias}, $^3$He clusters~\cite{barranco}, metal clusters~\cite{puente}, etc.\par
%
%------------------------------------------------------------------------------------------------
%
{}Furthermore, the PDM presence in quantum mechanical problems may reflect some other unconventional effects, such as a deformation of the canonical commutation relations or a curvature of the underlying space~\cite{cq04}. It has also recently been signalled in the rapidly growing field of PT-symmetric~\cite{bender98, bender04} (or pseudo-Hermitian~\cite{mosta04} or else quasi-Hermitian~\cite{scholtz}) quantum mechanics as occurring in the Hermitian Hamiltonian equivalent to some PT-symmetric systems at lowest order of perturbation theory~\cite{jones, mosta05, bagchi06a}.\par
%
%----------------------------------------------------------------------
%
Looking for exact solutions of the Schr\"odinger equation with a PDM has become an interesting research topic because such solutions may provide a conceptual understanding of some physical phenomena, as well as a testing ground for some approximation schemes. The generation of PDM and potential pairs leading to exactly solvable, quasi-exactly solvable or conditionally exactly solvable equations has been achieved by
extending some methods known in the constant-mass case, such as point canonical transformations~\cite{dekar98, dekar99,dutra00, gonul02a, gonul02b, alhaidari,bagchi04a, bagchi05a, bagchi06b, cq06, cq07a, yu04a, yu04b, chen, dong, jiang, mustafa06a, mustafa06b, ganguly, carinena}, Lie algebraic methods~\cite{roy02, roy05a, koc02, koc03a, bagchi04b}, and supersymmetric quantum mechanical techniques (or related intertwining operator methods)~\cite{cq04, gonul02a, gonul02b, bagchi04a, bagchi05a, cq06, carinena, roy02, milanovic, plastino, dutra03, roy05b, koc03b, gonul05, gonul06, bagchi05b, tanaka}.\par
%
%------------------------------------------------------------------------------------
%
Another powerful tool used in standard quantum mechanics is that of nonlinear algebras, more specifically quadratic ones. For one-dimensional systems allowing exact solutions, such algebras may help us to understand the relation between the time evolution of classical dynamical variables and that of corresponding quantum operators, while providing a general method for constructing spectrum generating algebras~\cite{granovskii92a} (see also \cite{odake}). In more than one dimension, they are a clue to classifying superintegrable systems with integrals of motion quadratic in the momenta~\cite{kalnins, daska06a, daska06b} and to solving the Schr\"odinger equation for such systems~\cite{granovskii92b, bonatsos, daska01}.\par
%
%------------------------------------------------------------------------------------
%
In a PDM context, there has been no systematic use of quadratic algebras so far, although the presence of one of them has been signalled in a one-dimensional problem~\cite{roy05b}. To start filling in this gap, we have recently considered the quadratic algebra generated by the integrals of motion of a two-dimensional superintegrable PDM system and shown how a deformed parafermionic oscillator realization of this algebra allows one to derive the bound-state energy spectrum~\cite{cq07b}.\par
%
%--------------------------------------------------------------------------------------------
%
In the present paper, we turn ourselves to another aspect of quadratic algebras, namely their occurrence as spectrum generating algebras, which we shall illustrate with the simplest example, corresponding to a harmonic oscillator potential. For a constant mass, it is well known (see, e.g.,~\cite{wybourne}) that all the states of such a potential with a given parity in one dimension or with a given angular momentum $l$ in more than one dimension belong to a single unitary irreducible representation of an su(1,1) Lie algebra. The corresponding lowest-energy state is annihilated by the lowering generator, while the remaining states can be obtained from it  by repeated applications of the raising generator. We plan to show that for a specific PDM choice, similar results apply except that su(1,1) gets deformed. We shall establish that a quadratic algebra approach provides us with a key to constructing such a deformed algebra, while allowing us at the same time to derive the bound-state energy spectrum.\par
%
%----------------------------------------------------------------------------------------------
%
In section 2, we present the Schr\"odinger equation of a PDM $d$-dimensional radial harmonic oscillator ($d \ge 2$) and review its bound-state energy spectrum. The corresponding spectrum generating algebra is constructed in section 3. In section 4, we show how the general $d$-dimensional results can be applied to the one-dimensional oscillator on  the full line with a similar PDM. Finally, section 5 contains the conclusion.\par
%
%==========================================================
%
\section{\boldmath Schr\"odinger equation of a PDM $d$-dimensional radial harmonic oscillator}

Whenever both the PDM $m(r)$ and the potential $V(r)$ only depend on the radial variable $r$, the corresponding $d$-dimensional Schr\"odinger equation is separable in spherical coordinates. On writing the radial wavefunction as $r^{-(d-1)/2} \psi(r)$, so that the normalization condition for $\psi(r)$ reads
\begin{equation}
  \int_0^\infty |\psi(r)|^2 dr = 1  \label{eq:HO-normal}
\end{equation}
we end up with the radial equation
\begin{equation}
  \left(- \frac{d}{dr} \frac{1}{M(r)} \frac{d}{dr} + \vteff(r)\right) \psi(r) = E \psi(r).  \label{eq:radial-PDM} 
\end{equation}
Here $M(r)$ is the dimensionless form of the mass function $m(r) = m_0 M(r)$, we have taken units wherein $\hbar = 2 m_0 = 1$ and 
\begin{equation*}
  \vteff(r) = \veff(r) - \frac{(d-1)M'}{2rM^2} + \frac{L(L+1)}{Mr^2}
\end{equation*}
where a prime denotes derivative with respect to $r$, $L$ is defined by $L = l + (d-3)/2$ in terms of the angular momentum quantum number $l$ and $\veff(r)$ is the effective potential that would arise in a cartesian coordinate approach to the problem (see equation (2.3) of \cite{cq06}).\par
%
%------------------------------------------------------------------------------------------------
%
Let us now consider a PDM $d$-dimensional harmonic oscillator, whose radial Schr\"odinger equation is obtained by replacing in the constant-mass one the radial momentum $p_r = - {\rm i} d/dr$ by some deformed operator, $\pi_r = \sqrt{f(\alpha; r)}\, p_r \sqrt{f(\alpha; r)}$, where $f(\alpha; r) = 1 + \alpha r^2$ and $\alpha$ is a positive real constant. The result of this substitution reads
\begin{equation}
  \left(\pi_r^2 + \frac{L(L+1)}{r^2} + \frac{1}{4} \omega^2 r^2\right) \psia(r) = \ea \psia(r)  
  \label{eq:PDM-HO}
\end{equation}
which is equivalent to (\ref{eq:radial-PDM}) with
\begin{equation*}
  M(\alpha; r) = \frac{1}{f^2(\alpha; r)} = \frac{1}{(1 + \alpha r^2)^2}
\end{equation*}
and
\begin{equation*}
  \vteff(r) = \frac{L(L+1)}{r^2} + \frac{1}{4} (\omega^2 - 8 \alpha^2) r^2 - \alpha
\end{equation*}
or
\begin{equation*}
  \veff(r) = \tfrac{1}{4} \{\omega^2 - 4\alpha^2 [L(L+1) + 2d]\} r^2 - \alpha [2L(L+1) + 2d -1].
\end{equation*}
Observe that the constant-mass limit corresponds to $\alpha \to 0$, in which case equation (\ref{eq:PDM-HO}) gives back the standard constant-mass equation.\par%
%-------------------------------------------------------------------------------------------------
%
Supersymmetric quantum mechanical methods, combined with deformed shape invariance, have shown~\cite{bagchi05b} that the PDM Schr\"odinger equation (\ref{eq:PDM-HO}) has an infinite number of bound states giving rise to a quadratic energy spectum
\begin{equation}
 \ea_{n,L} = \alpha \left(4n^2 + 4n(L+1) + L + 1 + (4n + 2L + 3) \frac{\lambda}{\alpha}\right) \qquad n=0, 1, 2, \ldots
 \label{eq:PDM-E}
\end{equation}
where $\lambda = \frac{1}{2}(\alpha + \Delta)$ and $\Delta = \sqrt{\omega^2 + \alpha^2}$. In the same work, the lowest-energy wavefunction (for given $L$) has been obtained in the form
\begin{equation}
  \psia_{0,L}(r) = \na_{0,L} r^{L+1} f^{-[\lambda + (L+2) \alpha]/(2\alpha)}  \label{eq:PDM-psi}
\end{equation}
where the normalization coefficient $\na_{0,L}$ can be easily determined from (\ref{eq:HO-normal}) as
\begin{equation*}
  \na_{0,L} = \left(\frac{2 \alpha^{L+\frac{3}{2}} \Gamma\left(\lambdaa + L + 2\right)}{
  \Gamma(L+\frac{3}{2}) \Gamma\left(\lambdaa + \frac{1}{2}\right)}\right)^{1/2}.
\end{equation*}
\par
%
%----------------------------------------------------------------------------------------------------------
%
Some lengthy calculations along the same lines also yield~\cite{cq05}
\begin{equation}
  \psia_{n,L}(r) = \frac{\na_{n,L}}{\na_{0,L}} P_n^{\left(\lambdaa - \frac{1}{2}, L + \frac{1}{2}\right)}(t)
  \psia_{0,L}(r)  \label{eq:PDM-psi-bis}
\end{equation}
where $P_n^{\left(\lambdaa - \frac{1}{2}, L + \frac{1}{2}\right)}(t)$ is a Jacobi polynomial~\cite{abramowitz} in the variable
\begin{equation}
  t = 1 - \frac{2}{f} = \frac{-1 + \alpha r^2}{1 + \alpha r^2}  \label{eq:t}
\end{equation}
and
\begin{equation}
  \frac{\na_{n,L}}{\na_{0,L}} = \left(\frac{\Gamma(L+\frac{3}{2}) \Gamma\left(\lambdaa + \frac{1}{2}\right)
  n! \left(\lambdaa + 2n+L+1\right) \Gamma\left(\lambdaa + n+L+1\right)}{\Gamma\left(\lambdaa + L+2  
  \right) \Gamma\left(\lambdaa + n +\frac{1}{2}\right) \Gamma\left(n + L + \frac{3}{2}\right)}
  \right)^{1/2}.  \label{eq:PDM-norm}   
\end{equation}
\par
%
%----------------------------------------------------------------------------------------------------
%
Since in the constant-mass limit, the parameter $\lambda$ goes over to $\omega/2$, it is clear that in such a limit the quadratic energy spectrum (\ref{eq:PDM-E}) becomes linear and given by $E_{n,L} = \omega \left(2n + L + \frac{3}{2}\right)$. Furthermore, the mere definition of $e$, combined with limit relations between orthogonal polynomials~\cite{abramowitz} also allows us to retrieve the results for constant-mass wavefunctions $\psi_{n,L}(r)$, depending on Laguerre polynomials~\cite{moshinsky}.\footnote{Note that we obtain a phase factor $(-1)^n$ not present in equation (28.5) of \cite{moshinsky}. This phase factor is consistent with positive matrix elements for the su(1,1) generators and with standard wavefunctions for the one-dimensional harmonic oscillator (see section 4).}\par
%
%========================================================
%
\section{\boldmath Spectrum generating algebra of the PDM $d$-dimensional radial harmonic oscillator}

In order to build a counterpart of the su(1,1) spectrum generating algebra obtained in the constant-mass case~\cite{wybourne}, it is useful to start from a quadratic algebra approach. It has indeed been suggested~\cite{granovskii92a, odake, granovskii92b} that for a whole class of Hamiltonians, such as those for which the bound-state wavefunctions can be written as the lowest-energy one multiplied by increasing-degree polynomials in some variable $t$, there may exist an (in general nonlinear) algebra generating the spectrum, whose three generators are the Hamiltonian and the variable $t$, which are Hermitian operators, as well as their anti-Hermitian commutator. This algebra is characterized by a Casimir operator, which is some polynomial function of the three generators~\cite{granovskii92a}. This is the approach to be followed in section 3.1.\par
%
%+++++++++++++++++++++++++++++++++++++++++++++++++++++
%
\subsection{Quadratic algebra approach to the spectrum generating algebra}

Let us start from the Hamiltonian defined in equation (\ref{eq:PDM-HO}), the variable $t$ considered in (\ref{eq:t}) and their commutator,
\begin{equation}
  \kta_1 = \pi_r^2 + \frac{L(L+1)}{r^2} + \frac{1}{4} \omega^2 r^2 \qquad \kta_2 = t \qquad \kta_3 = 
  - 4{\rm i}{\alpha} \left(2 \frac{r}{f} \pi_r + {\rm i}t\right).  \label{eq:PDM-gen}
\end{equation}
\par
%
%--------------------------------------------------------------------------------------------------
%
{}From the basic commutator $[r, \pi_r] = {\rm i} f(\alpha; r)$, it is straightforward to derive the relations
\begin{equation}
\begin{split}
  \bigl[\kta_1, \kta_2\bigr] &= \kta_3 \\
  \bigl[\kta_2, \kta_3\bigr] &= 8 \alpha \bigl(1 - \tilde{K}^{(\alpha)2}_2\bigr) \\
  \bigl[\kta_3, \kta_1\bigr] &= - 8 \alpha \bigl\{\kta_1, \kta_2\bigr\} - 16 \alpha^2 \left[\lambdaa 
    \left(\lambdaa - 1\right) + L(L+1) - 1\right] \kta_2 \\
  & \quad - 16 \alpha^2 \left[\lambdaa \left(\lambdaa - 1\right) - L(L+1)\right] 
\end{split}  \label{eq:PDM-com}
\end{equation}
showing that the operators $\kta_1$, $\kta_2$ and $\kta_3$ generate a quadratic algebra. Its nature can be determined  by comparing (\ref{eq:PDM-com}) with equation (3.2) of \cite{granovskii92a}, defining the (general) Askey-Wilson algebra QAW(3) in terms of eight parameters $R$, $A_1$, $A_2$, $C_1$, $C_2$, $D$, $G_1$ and $G_2$. Since in the present case, $R = A_1 = C_1 = 0$, we have to deal here with a quadratic Jacobi algebra QJ(3), characterized by the parameters
\begin{equation}
\begin{split}
  A_2 &= - 8 \alpha \qquad C_2 = - 16 \alpha^2 \left[\lambdaa \left(\lambdaa - 1\right) + L(L+1) - 1\right]
    \qquad D = 0 \qquad G_1 = 8 \alpha \\
  G_2 & = - 16 \alpha^2 \left[\lambdaa \left(\lambdaa - 1\right) - L(L+1)\right].   
\end{split}  \label{eq:Jacobi-para}
\end{equation}
As $D^2 - 4 A_2 G_1 \ne 0$, this algebra is a nondegenerate one, i.e., an algebra that cannot be reduced to a Lie algebra by a change of basis.\par
%
%------------------------------------------------------------------------------------------------
%
{}From equation (3.4) of \cite{granovskii92a}, we get the corresponding Casimir operator in the form
\begin{equation}
\begin{split}
  \qa &= - 16 \alpha \kta_2 \kta_1 \kta_2 + \tilde{K}^{(\alpha)2}_3 - 16 \alpha^2 \left[\lambdaa \left(\lambdaa   
    - 1\right) + L(L+1) - 1\right] \tilde{K}^{(\alpha)2}_2 \\
  & \quad + 16 \alpha \kta_1 - 32 \alpha^2 \left[\lambdaa \left(\lambdaa - 1\right) - L(L+1)\right] \kta_2.  \label{eq:PDM-Q}  
\end{split}
\end{equation}
Its eigenvalue can be obtained by inserting the explicit expressions (\ref{eq:PDM-gen}) in (\ref{eq:PDM-Q}) and is given by
\begin{equation}
  \qa = 16 \alpha^2 \left[\lambdaa \left(\lambdaa - 1\right) + L(L+1) - 2\right].  \label{eq:PDM-Q-bis}
\end{equation}
\par
%
%---------------------------------------------------------------------------------------------------------
%
Our aim now consists in constructing a positive-discrete series unitary irreducible representation of this algebra spanned by the Hamiltonian eigenfunctions $\psia_{n,L}(r)$, $n=0$, 1, 2,~\ldots, which will be a counterpart of the su(1,1) representation $D^+_k$ with $k = \frac{1}{2} \left(L + \frac{3}{2}\right)$, obtained in the constant-mass case~\cite{wybourne}.\par
%
%-----------------------------------------------------------------------------------------------------------
%
{}From the general theory developed in \cite{granovskii92a, granovskii92b}, we know that in a basis $\psi_p$ wherein the Hamiltonian, i.e., the generator $\kta_1$, is diagonal, the unitary irreducible representations of QJ(3) are given by
\begin{equation*}
\begin{split}
  \kta_1 \psi_p &= \lambda_p \psi_p \\
  \kta_2 \psi_p &= a_{p+1} \psi_{p+1} + a_p \psi_{p-1} + b_p \psi_p \\
  \kta_3 \psi_p &= g_{p+1} a_{p+1} \psi_{p+1} - g_p a_p \psi_{p-1}
  \end{split}
\end{equation*}
where $\lambda_p$, $a_p$, $b_p$ and $g_p$ are some real constants, which can be expressed in terms of the defining parameters (\ref{eq:Jacobi-para}) and read
\begin{equation}
\begin{split}
  \lambda_p &= \alpha \left[4p(p+1) - \lambdaa(\lambdaa-1) - L(L+1) + 1\right] \\
  a_p^2 &= [16p^2(2p-1)(2p+1)]^{-1} \left(2p - \lambdaa + L + 1\right) \left(2p - \lambdaa - L\right) \\
  & \quad \times \left(2p + \lambdaa - L - 1\right) \left(2p + \lambdaa + L\right) \\
  b_p &= - [4p(p+1)]^{-1} \left(\lambdaa - L - 1\right) \left(\lambdaa + L\right) \\
  g_p &= 8 \alpha p.
\end{split} \label{eq:irrep-para}
\end{equation}
\par
%
%-------------------------------------------------------------------------------------------------------
%
An infinite-dimensional representation of the positive-discrete series type $D^+_{p_0}$ is then characterized by the properties $a_{p_0}^2 = 0$ and $a_p^2 > 0$ if $p = p_0 + n$, $n=1$, 2,~\ldots. From the explicit value of $a_p^2$ given in (\ref{eq:irrep-para}), it is clear that, for generic values of $\lambda/\alpha$ and $L$, such conditions can be achieved in a single way, namely by assuming
\begin{equation}
  p_0 = \frac{1}{2}\left(\lambdaa + L\right).  \label{eq:p-0}
\end{equation}
From (\ref{eq:irrep-para}) and (\ref{eq:p-0}), it results that the eigenvalues $\lambda_{p_0+n}$ of $\kta_1$ in $D^+_{p_0}$ coincide with the energy eigenvalues (\ref{eq:PDM-E}), i.e., $\lambda_{p_0+n} = \ea_{n,L}$, $n=0$, 1, 2,~\ldots.\par
%
%-----------------------------------------------------------------------------------------------------------------------
%
{}Furthermore, if we reset $\psi_{p_0+n} \to \psia_{n,L}$, $a_{p_0+n} \to \aal_{n,L}$, $b_{p_0+n} \to \ba_{n,L}$ and $g_{p_0+n} \to \ga_{n,L}$, the action of the generators $\kta_2$ and $\kta_3$ on the basis functions can be recast in the form
\begin{equation}
\begin{split}
  \kta_2 \psia_{n,L} &= \aal_{n+1,L} \psia_{n+1,L} + \aal_{n,L} \psia_{n-1,L} + \ba_{n,L} \psia_{n,L} \\
  \kta_3 \psia_{n,L} &= \ga_{n+1,L} \aal_{n+1,L} \psia_{n+1,L} - \ga_{n,L} \aal_{n,L} \psia_{n-1,L}
\end{split}  \label{eq:gen-action}
\end{equation}
where
\begin{equation}
\begin{split}
  \aal_{n,L} &= \frac{\tau_n}{\lambdaa+2n+L} \left(\frac{n (2n+2L+1) \left(\lambdaa+n+L\right) \left(2\lambdaa
    +2n-1\right)}{\left(\lambdaa+2n+L-1\right) \left(\lambdaa+2n+L+1\right)}\right)^{1/2} \\
  \ba_{n,L} &= - \frac{\left(\lambdaa-L-1\right) \left(\lambdaa+L\right)}{\left(\lambdaa+2n+L\right) 
    \left(\lambdaa+2n+L+2\right)} \\
  \ga_{n,L} &= 4\alpha \left(\lambdaa+2n+L\right)  
\end{split} \label{eq:a-b-g}
\end{equation}
and $\tau_n$ is a phase factor depending on the choice made for the relative phase of $\psia_{n,L}$ and $\psia_{n-1,L}$. The first equation in (\ref{eq:gen-action}) can be reduced to the recursion relation for the Jacobi polynomials $P_n^{\left(\lambdaa - \frac{1}{2}, L + \frac{1}{2}\right)}(t)$ and with the choice made in (\ref{eq:PDM-norm}) for the normalization coefficients, we find that $\tau_n = +1$.\par
%
%------------------------------------------------------------------------------------------------------------
%
We conclude that the solutions of the PDM Schr\"odinger equation (\ref{eq:PDM-HO}) can be derived by only using the quadratic algebra generated by the operators (\ref{eq:PDM-gen}). To obtain from the latter the generators of a deformed su(1,1) spectrum generating algebra (and consequently a simpler construction of wavefunctions), we shall need to build some ladder operators, generalizing the constant-mass ones. Before proceeding to such a derivation in section 3.3, it is worth considering the constant-mass limit of the quadratic algebra that we have just introduced.\par
%
%+++++++++++++++++++++++++++++++++++++++++++++++++++++++++++
%
\subsection{Constant-mass limit of the quadratic algebra}

Although appropriate for solving the Schr\"odinger equation (\ref{eq:PDM-HO}), the basis $\bigl(\kta_1, \kta_2, \kta_3\bigr)$ of our quadratic algebra is not convenient to determine its $\alpha \to 0$ limit because $\kta_2$ goes over to the constant $-1$. To circumvent this difficulty, it is necessary to go from $\bigl(\kta_1, \kta_2, \kta_3\bigr)$ to a new basis $\bigl(\kba_1, \kba_2, \kba_3\bigr)$.\par
%
%----------------------------------------------------------------------------------------------------
% 
Let us set
\begin{equation*}
\begin{split}
  \kba_1 & = \kta_1 = \pi_r^2 + \frac{L(L+1)}{r^2} + \frac{1}{4} \omega^2 r^2 \\
  \kba_2 & = \frac{1} {\alpha} \bigl(1 - \kta_2\bigr)^{-1} \bigl(1 + \kta_2\bigr) = r^2 \\ 
  \kba_3 & = \frac{1}{2\alpha} \bigl\{\bigl(1 - \kta_2\bigr)^{-1}, \kta_3\bigr\} = - 2 (2 {\rm i} r \pi_r + f).
\end{split}
\end{equation*}
Observe that the inverse transformation reads
\begin{equation}
  \kta_1 = \kba_1 \qquad \kta_2 = \bigl(1 + \alpha \kba_2\bigr)^{-1} \bigl(- 1 + \alpha \kba_2\bigr) \qquad
  \kta_3 = \alpha \bigl\{\bigl(1 + \alpha \kba_2\bigr)^{-1}, \kba_3\bigr\}.  \label{eq:gen-transf}
\end{equation}
\par
%
%-----------------------------------------------------------------------------------------------------
%
Either from the commutation relations (\ref{eq:PDM-com}) of the first basis generators or by direct computation, we obtain for the second basis the commutation relations
\begin{equation*}
\begin{split}
  \bigl[\kba_1, \kba_2\bigr] &= \tfrac{1}{2} \bigl\{1 + \alpha \kba_2, \kba_3\bigr\} \\
  \bigl[\kba_2, \kba_3\bigr] &= 8 \kba_2 \bigl(1 + \alpha \kba_2\bigr) \\
  \bigl[\kba_3, \kba_1\bigr] &= 4 \bigl\{1 + \alpha \kba_2, \kba_1\bigr\} - 16 \alpha^2 \lambdaa 
    \left(\lambdaa-1\right) \kba_2 \bigl(1 + \alpha \kba_2\bigr) \\
  & \quad + 4 \alpha \bigl(1 + \alpha \kba_2\bigr) \bigl(1 + 3\alpha \kba_2\bigr). 
\end{split}
\end{equation*}
In the $\alpha \to 0$ limit, it is obvious that these relations become linear. It is then straightforward to show that the resulting operators $\kb_i = \lim_{\alpha \to 0} \kba_i$, $i=1, 2, 3$, are some linear combinations of su(1,1) generators $K_0$, $K_+$, $K_-$, with commutation relations $[K_0, K_\pm] = \pm K_{\pm}$ and $[K_+, K_-] = - 2 K_0$. The results read $\kb_1 = 2 \omega K_0$, $\kb_2 = (2/\omega) (K_+ + K_- + 2 K_0)$ and $\kb_3 = 4 (K_+ - K_-)$.\par
%
%---------------------------------------------------------------------------------------------------------
%
{}Finally, on performing transformation (\ref{eq:gen-transf}) on the right-hand side of (\ref{eq:PDM-Q}), the quadratic algebra Casimir operator yields, after some calculations, the relation
\begin{equation}
\begin{split}
  &\quad \qa - 16 \alpha^2 \left[\lambdaa \left(\lambdaa-1\right) + L(L+1) - 2\right] \\
  & \quad = \bigl(1 + \alpha \kba_2\bigr)^{-1} \biggl\{4 \alpha^2 \biggl[\bar{K}^{(\alpha)2}_3 - 16 \alpha^2 
    \lambdaa \left(\lambdaa-1\right) \bar{K}^{(\alpha)2}_2 + 8 \bigl\{\kba_1, \kba_2\bigr\} + 12 \\
  & \qquad  - 16 L(L+1)\biggr] + 160 \alpha^3 \kba_2 + 112 \alpha^4 \bar{K}^{(\alpha)2}_2\biggr\} 
    \bigl(1 + \alpha \kba_2\bigr)^{-1}. 
\end{split}  \label{eq:PDM-Q-ter}
\end{equation}
From equation (\ref{eq:PDM-Q-bis}), it follows that the operator between curly brackets on the right-hand side of (\ref{eq:PDM-Q-ter}) vanishes. Since $\omega^2 = 4 \alpha^2 \lambdaa \left(\lambdaa-1\right)$, we observe a close similarity between the first few terms making up this operator and the expression of the su(1,1) Casimir operator $C = - K_+ K_- + K_0 (K_0 - 1)$ in terms of $\kb_1$, $\kb_2$, $\kb_3$, namely $C = \left(\kb_3^2 - 4 \omega^2 \kb_2^2 + 8 \{\kb_1, \kb_2\}\right)/64$. We conclude that the substitution of a PDM for a constant mass has the effect  of changing the constant $C = \frac{1}{4} \left(L + \frac{3}{2}\right) \left(L - \frac{1}{2}\right)$ into a function of $r$,
\begin{equation}
\begin{split}
  \bar{C}^{\alpha}(r) &\equiv \frac{1}{64} \left[\bar{K}^{(\alpha)2}_3 - 16 \alpha^2 \lambdaa \left
    (\lambdaa-1\right) \bar{K}^{(\alpha)2}_2 + 8 \bigl\{\kba_1, \kba_2\bigr\}\right] \\ 
  &= \frac{1}{16} \left[(2L+3) (2L-1) - 10 \alpha r^2 - 7 \alpha^2 r^4\right].
\end{split}  \label{eq:PDM-C}
\end{equation}
\par
%
%+++++++++++++++++++++++++++++++++++++++++++++++++++++++++++++++++
\subsection{Deformed su(1,1) spectrum generating algebra}

The purpose of this subsection is to construct a third basis $\bigl(\ka_0, \ka_+, \ka_-\bigr)$ of our quadratic algebra, satisfying the following three properties:
\begin{itemize}
\item[(i)] $\ka_0$ is proportional to the Hamiltonian  of the problem, while $\ka_+$ (resp.\ $\ka_-$) is a raising (resp.\ lowering) ladder operator, which means that, up to some multiplicative factor, it transforms $\psia_{n,L}$ into $\psia_{n+1,L}$ (resp.\ $\psia_{n-1,L}$) for any $n \in \N$ (resp.\ $n \in \N^+$) with the additional condition that $\ka_-$ annihilates $\psia_{0,L}$.
\item[(ii)] The operators $\ka_0$, $\ka_+$, $\ka_-$ satisfy the same Hermiticity properties as $K_0$, $K_+$, $K_-$, i.e., $K^{(\alpha)\dagger}_0 = \ka_0$ and $K^{(\alpha)\dagger}_{\pm} = \ka_{\mp}$.
\item[(iii)] In the $\alpha \to 0$ limit, they go over to the su(1,1) generators $K_0$, $K_+$, $K_-$.
\end{itemize}
\par
%
%------------------------------------------------------------------------------------------------------------
%
{}From the known action of $\kta_2$ and $\kta_3$ on $\psia_{n,L}$, given in (\ref{eq:gen-action}), we can construct some $n$-dependent ladder operators
\begin{equation}
  \Aa_{+,n} = \kta_3 + \ga_{n,L} \kta_2 - \ga_{n,L} \ba_{n,L} \qquad \Aa_{-,n} = \kta_3 - \ga_{n+1,L} \kta_2   
  + \ga_{n+1,L} \ba_{n,L}.  \label{eq:A-n} 
\end{equation}
It is indeed easy to check that
\begin{equation*}
  \Aa_{+,n} \psia_{n,L} = \aal_{n+1,L} \bigl(\ga_{n,L} + \ga_{n+1,L}\bigr) \psia_{n+1,L} \qquad
  \Aa_{-,n} \psia_{n,L} = - \aal_{n,L} \bigl(\ga_{n,L} + \ga_{n+1,L}\bigr) \psia_{n-1,L}.
\end{equation*}
In (\ref{eq:A-n}), the quantum number $n$ can be expressed in terms of $\ea_{n,L}$ by inverting equation (\ref{eq:PDM-E}) and choosing the nonnegative root of the resulting quadratic equation. The result reads
\begin{equation*}
  n = \frac{1}{2} \left[- \left(\lambdaa+L+1\right) + \delta_n\right] \qquad \delta_n = \sqrt{\frac{\ea_{n,L}} 
  {\alpha} + \lambdaa \left(\lambdaa-1\right) + L(L+1)}.
\end{equation*}
\par
%
%------------------------------------------------------------------------------------------------------
%
We can now eliminate the $n$ dependence from $\Aa_{\pm,n}$ by replacing $\ea_{n,L}$ by the Hamiltonian $H = \kta_1$. This leads to the operators
\begin{equation}
  \Aa_{\pm} = \kta_3 - 4\alpha \kta_2 (1 \mp \delta) + 4\alpha \frac{\left(\lambdaa-L-1\right) (\left(\lambdaa
  +L\right)}{1 \pm \delta}  \label{eq:A}
\end{equation}
where
\begin{equation}
  \delta = \sqrt{\frac{\kta_1}{\alpha} + \lambdaa \left(\lambdaa-1\right) + L(L+1)}.  \label{eq:delta}
\end{equation}
Although such operators satisfy condition (i) referred to above, they do not fulfil the remaining two conditions.\par
%
%-----------------------------------------------------------------------------------------------------------
%
We can get rid of this shortcoming by multiplying $\Aa_{\pm}$ by some appropriate functions $F^{(\alpha)}_{\pm}\bigl(\kta_1\bigr)$ of the Hamiltonian. Since the latter are not univoquely determined by conditions (ii) and (iii), we may choose them in such a way that the action of $\ka_{\pm}$ on $\psia_{n,L}$ is the simplest possible. Let us therefore define
\begin{equation}
  \ka_{\pm} = \pm \frac{1}{16\lambda} \Aa_{\pm} (\delta \pm 1) \sqrt{\frac{\delta \pm 2}{\delta}} = \pm 
  \frac{1}{16\lambda} (\delta \mp 1) \sqrt{\frac{\delta}{\delta \mp 2}} \Aa_{\pm}  \label{eq:PDM-gen-bis}  
\end{equation}
leading to the relations
\begin{equation}
\begin{split}
  \ka_+ \psia_{n,L} &= \frac{\alpha}{\lambda} \left[(n+1) \left(n+L+\frac{3}{2}\right) \left(n+\lambdaa+L+1  
    \right) \left(n+\lambdaa+\frac{1}{2}\right)\right]^{1/2} \psia_{n+1,L} \\
  \ka_- \psia_{n,L} &= \frac{\alpha}{\lambda} \left[n \left(n+L+\frac{1}{2}\right) \left(n+\lambdaa+L\right) 
    \left(n+\lambdaa-\frac{1}{2}\right)\right]^{1/2} \psia_{n-1,L}.
\end{split}  \label{eq:gen-action-bis}
\end{equation}
In (\ref{eq:PDM-gen-bis}), the factors $\pm \sqrt{(\delta \pm 2)/\delta}$ (alternatively $\pm \sqrt{\delta/(\delta\mp 2)}$) are required by condition (ii) above, whereas the factors $(\delta \pm 1)$ (alternatively $(\delta \mp 1)$) are optional ones having a simplifying effect on the matrix elements contained in (\ref{eq:gen-action-bis}).\par
%
%---------------------------------------------------------------------------------------------------------
%
The definition of the third basis is finally completed by
\begin{equation*}
  \ka_0 = \frac{1}{4\lambda} \kta_1
\end{equation*}
such that
\begin{equation}
  \ka_0 \psia_{n,L} = \frac{1}{4\lambda} \ea_{n,L} \psia_{n,L}.  \label{eq:gen-action-ter}
\end{equation}
In the $\alpha \to 0$ limit, equations (\ref{eq:gen-action-bis}) and (\ref{eq:gen-action-ter}) agree with the standard su(1,1) results $K_{\pm} \psi_{n,L}(r) = \left[\left(n + \frac{1}{2} \pm \frac{1}{2}\right) \left(n + L + 1 \pm \frac{1}{2}\right)\right]^{1/2} \psi_{n\pm1, L}(r)$ and $K_0 \psi_{n,L}(r) = (E_{n,L}/2\omega) \psi_{n,L}(r)$, respectively.\par
%
%------------------------------------------------------------------------------------------------------
%
The three deformed su(1,1) generators $\ka_0$, $\ka_+$ and $\ka_-$ satisfy the commutation relations
\begin{equation*}
  \bigl[\ka_0, \ka_{\pm}\bigr] = \pm \frac{\alpha}{\lambda} \ka_{\pm} (\delta \pm 1) = \pm 
  \frac{\alpha}{\lambda} (\delta \mp 1) \ka_{\pm} \qquad \bigl[\ka_+, \ka_-\bigr] = 
  - \frac{\alpha\delta}{\lambda} \left(2 \ka_0 + \frac{\alpha}{4\lambda}\right)   
\end{equation*}
which can be easily checked by applying both sides on any $\psia_{n,L}$. Observe that for $\alpha \to 0$, we get $\alpha\delta/\lambda \to 1$ and $\alpha/\lambda \to 0$, so that standard su(1,1) commutation relations are retrieved, as it should be.\par
%
%-----------------------------------------------------------------------------------------------------
%
The Casimir operator $\ca$ of this deformed su(1,1) algebra can be written as $\ca = - \ka_+ \ka_- + f\bigl(\ka_0\bigr)$, where the function $f\bigl(\ka_0\bigr)$ must be such that $\ca$ commutes with $\ka_+$ and that $f\bigl(\ka_0\bigr) \to K_0(K_0-1)$ for $\alpha \to 0$. The latter condition of course determines $\ca$ only up to some constant term of order $O(\alpha/\lambda)$. After some rather lengthy calculations, we arrive at the result
\begin{equation*}
  \ca = - \ka_+ \ka_- + K^{(\alpha)2}_0 - \frac{\alpha}{\lambda} \left(\delta - \frac{5}{4}\right) \ka_0 -
  \frac{\alpha^2}{8\lambda^2} \delta 
\end{equation*}
leading to
\begin{equation}
  \ca \psia_{n,L} = \left[\frac{1}{4} \left(1 - \frac{\alpha}{\lambda}\right) \left(L + \frac{3}{2}\right)
  \left(L - \frac{1}{2}\right) - \frac{3\alpha^2}{16\lambda^2} L(L+1)\right] \psia_{n,L}.  
  \label{eq:PDM-C-bis}
\end{equation}
Equation (\ref{eq:PDM-C-bis}) should be contrasted with (\ref{eq:PDM-C}).\par
%
%----------------------------------------------------------------------------------------------------
%
In the appendix, it is shown how the ladder operators $\ka_+$ and $\ka_-$ can be used to fully determine the functions $\psia_{n,L}$ in a much more direct way than those sketched above equation (\ref{eq:PDM-psi-bis}) and below equation (\ref{eq:a-b-g}).\par
%
%==============================================================
%
\section{One-dimensional harmonic oscillator case}

The purpose of this section is to show how the results of section 3, valid for $d \ge 2$, can be extended to the one-dimensional harmonic oscillator on the full line. This implies, in particular, replacing the radial variable $r$ ($0 < r < \infty$) by $x$ ($-\infty < x < \infty$).\par
%
%--------------------------------------------------------------------------------------------------
% 
{}For a constant mass, it is well known that apart from the substitution $r \to x$, the Schr\"odinger equation for the standard one-dimensional harmonic oscillator can be deduced from the $d$-dimensional radial one by setting either $L=-1$ or $L=0$. In the former (resp.\ latter) case, one gets the even-parity (resp.\ odd-parity) wavefunctions and corresponding eigenvalues, $\psi_{\nu, -1}(r)/\sqrt{2} \to \psi_{2\nu}(x)$, $E_{\nu, -1} \to E_{2\nu}$ (resp.\ $\psi_{\nu, 0}(r)/\sqrt{2} \to \psi_{2\nu+1}(x)$, $E_{\nu, 0} \to E_{2\nu+1}$), due to some relations between Laguerre and Hermite polynomials~\cite{abramowitz}. As a consequence, the single su(1,1) unitary irreducible representation $D^+_k$, $k = \frac{1}{2} \left(L + \frac{3}{2}\right)$, of the radial case gives rise to two such representations $D^+_{1/4}$ and $D^+_{3/4}$ (with the same Casimir $C = - 3/16$), for the one-dimensional case.\par
%
%--------------------------------------------------------------------------------------------------
%
The PDM Schr\"odinger equation
\begin{gather*}
  \bigl(\pi^2 + \tfrac{1}{4} \omega^2 x^2\bigr) \psia(x) = \ea \psia(x) \\
  \pi = \sqrt{f(\alpha;x)}\, p \sqrt{f(\alpha;x)} \qquad p = - {\rm i} \frac{d}{dx} \qquad f(\alpha;x) = 1 +  
    \alpha x^2
\end{gather*}
equivalent to 
\begin{gather*}
  \left(- \frac{d}{dx} \frac{1}{M(x)} \frac{d}{dx} + \veff(x)\right) \psia(x) = \ea \psia(x) \\
  M(x) = \frac{1}{f^2(\alpha;x)} = \frac{1}{(1 + \alpha x^2)^2} \qquad \veff(x) = \frac{1}{4} (\omega^2 -
    8 \alpha^2) x^2 - \alpha
\end{gather*} 
admits a similar treatment exploiting the results obtained for equation (\ref{eq:PDM-HO}), provided we distinguish again between the even- and odd-parity wavefunctions, given by
\begin{equation*}
  \psia_{2\nu}(x) = \frac{\na_{2\nu}}{\na_0} P_{\nu}^{\left(\lambdaa - \frac{1}{2}, - \frac{1}{2}\right)}(t)
  \psia_0(x) \qquad \psia_0(x) = \na_0 f^{-(\lambda+\alpha)/(2\alpha)}
\end{equation*}
and
\begin{equation*}
  \psia_{2\nu+1}(x) = \frac{\na_{2\nu+1}}{\na_1} P_{\nu}^{\left(\lambdaa - \frac{1}{2}, \frac{1}{2}\right)}
  (t) \psia_1(x) \qquad \psia_1(x) = \na_1 x\, f^{-(\lambda+2\alpha)/(2\alpha)}
\end{equation*}
respectively. Here $\nu=0$, 1, 2,~\ldots, $t = 1 - (2/f) = (-1 + \alpha x^2)/(1 + \alpha x^2)$,
\begin{align*}
  \na_0 &= \left(\frac{\sqrt{\alpha}\, \Gamma\left(\frac{\lambda}{\alpha}+1\right)}{\sqrt{\pi}\, \Gamma\left
    (\frac{\lambda}{\alpha}+\frac{1}{2}\right)}\right)^{1/2} & \frac{\na_{2\nu}}{\na_0} &= \left(\frac{\sqrt{\pi}\,
    \Gamma\left(\frac{\lambda}{\alpha}+\frac{1}{2}\right) \nu!\, \left(\frac{\lambda}{\alpha}+2\nu\right) 
    \Gamma\left(\frac{\lambda}{\alpha}+\nu\right)}{\Gamma\left(\frac{\lambda}{\alpha}+1\right) \Gamma\left   
    (\frac{\lambda}{\alpha}+\nu+\frac{1}{2}\right) \Gamma\left(\nu+\frac{1}{2}\right)}\right)^{1/2} \\
  \na_1 &= \left(\frac{2\alpha^{3/2} \Gamma\left(\frac{\lambda}{\alpha}+2\right)}{\sqrt{\pi}\, \Gamma\left
    (\frac{\lambda}{\alpha}+\frac{1}{2}\right)}\right)^{1/2} & \frac{\na_{2\nu+1}}{\na_1} &= \left(\frac{\sqrt{\pi}
    \, \Gamma\left(\frac{\lambda}{\alpha}+\frac{1}{2}\right) \nu!\, \left(\frac{\lambda}{\alpha}+2\nu+1\right) 
    \Gamma\left(\frac{\lambda}{\alpha}+\nu+1\right)}{2\, \Gamma\left(\frac{\lambda}{\alpha}+2\right) \Gamma  
    \left (\frac{\lambda}{\alpha}+\nu+\frac{1}{2}\right) \Gamma\left(\nu+\frac{3}{2}\right)}\right)^{1/2}    
\end{align*}
and the corresponding eigenvalues are
\begin{equation*}
  \ea_n = \alpha \left(n^2 + (2n+1) \lambdaa\right) \qquad \lambda = \frac{1}{2} (\alpha + \Delta) \qquad
  \Delta = \sqrt{\omega^2 + \alpha^2}
\end{equation*}
in both cases $n = 2\nu$ and $n = 2\nu+1$.\par
%
%-----------------------------------------------------------------------------------------------------
%
There exists a quadratic spectrum generating algebra, for which we can construct three sets of generators $\bigl(\kta_1, \kta_2, \kta_3\bigr)$, $\bigl(\kba_1, \kba_2, \kba_3\bigr)$ and $\bigl(\ka_0, \ka_+, \ka_-\bigr)$, analogous to those built in section 3. The only differences lie in the substitutions $r \to x$, $\pi_r \to \pi$, $L(L+1) \to 0$, and in the very important fact that there are now two distinct unitary irreducible representations instead of a single one. This can be seen from the counterpart
\begin{equation*}
  a_p^2 = [16 p^2 (2p-1) (2p+1)]^{-1} \left(2p - \lambdaa\right) \left(2p - \lambdaa + 1\right)
  \left(2p + \lambdaa\right) \left(2p + \lambdaa - 1\right)
\end{equation*}
of the similar quantity defined in (\ref{eq:irrep-para}). The conditions $a_{p_0}^2 = 0$ and $a_p^2 > 0$ if $p = p_0 + \nu$, $\nu=1$, 2,~\ldots, characterizing positive-discrete series representations $D^+_{p_0}$, are indeed satisfied now by two distinct values of $p_0$, $p_0 = \frac{1}{2} \left(\lambdaa - 1\right)$ and $p_0 = \frac{\lambda}{2\alpha}$, corresponding to $L = -1$ and $L = 0$ in (\ref{eq:p-0}) and to which we can associate $\lambda_{p_0+\nu} = \ea_{2\nu}$ and $\lambda_{p_0+\nu} = \ea_{2\nu+1}$, respectively.\par
%
%-------------------------------------------------------------------------------------------------
%
Since, after these observations, it is straightforward to transpose the results of section 3 to the one-dimensional case, we are not going to detail them here. We would only like to mention that the action of the deformed su(1,1) generators on the wavefunctions reads
\begin{equation*}
\begin{split}
  \ka_0 \psia_n(x) &= \frac{1}{4\lambda} \ea_n \psia_n(x) = \frac{\alpha}{4\lambda} \left(n^2 + (2n+1)
    \lambdaa\right) \psia_n(x) \\
  \ka_+ \psia_n(x) &= \frac{\alpha}{4\lambda} \left[(n+1) (n+2) \left(n + 2\lambdaa\right) \left(n + 2   
    \lambdaa + 1\right)\right]^{1/2} \psia_{n+2}(x) \\
  \ka_- \psia_n(x) &= \frac{\alpha}{4\lambda} \left[n (n-1) \left(n + 2\lambdaa - 2\right) \left(n + 2   
    \lambdaa - 1\right)\right]^{1/2} \psia_{n-2}(x) 
\end{split}
\end{equation*}
leading to the standard su(1,1) results $K_0 \psi_n(x) = \frac{1}{2} \left(n + \frac{1}{2}\right) \psi_n(x)$, $K_{\pm} \psi_n(x) = \frac{1}{2} [(n \pm 1) (n + 1 \pm 1)]^{1/2} \psi_{n \pm 2}(x)$ in the $\alpha \to 0$ limit.\par
%
%==================================================================
% 
\section{Conclusion}

In this paper, we have highlighted the interest of quadratic algebras for PDM Schr\"odinger equations by constructing spectrum generating algebras for a class of $d$-dimensional radial harmonic oscillators with $d \ge 2$ and a specific PDM choice, depending on some positive parameter $\alpha$. We have also shown how minor changes enable the one-dimensional oscillator on the line with the same type of mass to be included in such a class.\par
%
%---------------------------------------------------------------------------------------------------------
%
{}For these quadratic algebras, we have considered three different sets of generators. The first one $\bigl(\kta_1, \kta_2, \kta_3\bigr)$ has allowed us to prove the existence of a single unitary irreducible representation belonging to the positive-discrete series type for $d \ge 2$ and of two of them for $d=1$, as well as to obtain the bound-state quadratic energy spectrum.\par
%
%-------------------------------------------------------------------------------------------------------
%
The second set $\bigl(\kba_1, \kba_2, \kba_3\bigr)$ has provided us with an explicit demonstration that the quadratic algebra considered here gives rise to the well-known su(1,1) Lie algebra generating the oscillator spectrum in the constant-mass limit, i.e., for $\alpha \to 0$.\par
%
%-------------------------------------------------------------------------------------------------
% 
This correspondence has been studied further by constructing a third set of operators $\bigl(\ka_0, \ka_+, \ka_-\bigr)$, which go over to the standard su(1,1) generators $(K_0, K_+, K_-)$ for $\alpha \to 0$ and may therefore be termed deformed su(1,1) generators. All the bound-state wavefunctions have finally been built by using the lowering and raising generators, $\ka_-$ and $\ka_+$, respectively.\par
%
%---------------------------------------------------------------------------------------------------------
%
Some interesting open problems for future work are the extensions of the present study to other exactly solvable PDM Schr\"odinger equations either with the same potential but a different mass or with both different potential and mass.\par
%
%===============================================================
%
\section*{Appendix}
\renewcommand{\theequation}{A.\arabic{equation}}

The purpose of this appendix is to prove equations (\ref{eq:PDM-psi})--(\ref{eq:PDM-norm}) by using the deformed su(1,1) algebra introduced in section 3.3.\par
%
%------------------------------------------------------------------------------------------------
%
Let us start with $\psia_{0,L}(r)$, which, according to the second relation in (\ref{eq:gen-action-bis}), is annihilated by $\ka_-$ or, equivalently, by $\Aa_-$. Equations (\ref{eq:A}) and (\ref{eq:delta}), together with (\ref{eq:PDM-gen}), yield the first-order differential equation
\begin{equation*}
  r \frac{d}{dr} \psia_{0,L}(r) = \left[- \frac{1}{2} \left(\lambdaa + 1\right) (1+t) + \frac{1}{2} (L+1) (1-t)
  \right] \psia_{0,L}(r)
\end{equation*}
whose solution can be written in the form (\ref{eq:PDM-psi}).\par
%
%-------------------------------------------------------------------------------------------------------
%
The excited-state wavefunctions $\psia_{n,L}(r)$, $n=1$, 2,~\ldots, can now be determined recursively from $\psia_{0,L}(r)$ by employing the first relation in (\ref{eq:gen-action-bis}). When combined with definition (\ref{eq:PDM-gen-bis}), the latter yields
\begin{equation}
\begin{split}
  \psia_{n+1,L}(r) &= \frac{1}{16\alpha} \left(2n + \lambdaa + L + 2\right) \left(2n + \lambdaa + L +  3
    \right)^{1/2} \\
  & \quad \times \left[(n+1) \left(n + L + \frac{3}{2} \right) \left(n + \lambdaa + L + 1\right) \left(n + 
    \lambdaa + \frac{1}{2}\right)\right]^{-1/2} \\
  & \quad \times \left(2n + \lambdaa + L + 1\right)^{-1/2} \Aa_+ \psia_{n,L}(r).  
\end{split} \label{eq:result1}  
\end{equation}
\par
%
%---------------------------------------------------------------------------------------------------
%
Let us now make the ansatz
\begin{equation}
  \psia_{n,L}(r) = \frac{\na_{n,L}}{\na_{0,L}} \psia_{0,L}(r) P_n(t)  \label{eq:result2}
\end{equation}
where $P_n(t)$ is some $n$th-degree polynomial in $t$, such that $P_0(t) = 1$. On inserting (\ref{eq:result2}) in $\Aa_+ \psia_{n,L}(r)$ and using equations (\ref{eq:PDM-gen}) and (\ref{eq:A}), we get
\begin{equation*}
\begin{split}
  \Aa_+ \psia_{n,L}(r) &= - 8\alpha \frac{\na_{n,L}}{\na_{0,L}} \frac{\psia_{0,L}(r)}{2n + \lambdaa + L +
    2} \biggl\{\left(2n + \lambdaa + L +2\right) (1 - t^2) \frac{d}{dt} \\
  & \quad - \left(n + \lambdaa + L +1\right) \left[\lambdaa - L - 1 + \left(2n + \lambdaa + L + 2\right) t
    \right] \biggr\} P_n(t) 
\end{split}
\end{equation*}
which, according to (\ref{eq:result1}) and (\ref{eq:result2}), should be proportional to $\psia_{0,L}(r) P_{n+1}(t)$. This clearly identifies $P_n(t)$ as the Jacobi polynomial $P_n^{(\beta,\gamma)}(t)$ with $\beta = \lambdaa - \frac{1}{2}$, $\gamma = L + \frac{1}{2}$, because the latter satisfies the relation
\begin{equation}
\begin{split}
  &\left\{(2n + \beta + \gamma + 2) (1 - t^2) \frac{d}{dt} - (n + \beta + \gamma + 1) [\beta - \gamma
    + (2n + \beta + \gamma + 2) t]\right\}  \\
  & \times P_n^{(\beta,\gamma)}(t) = - 2 (n+1) (n + \beta + \gamma + 1) P_{n+1}^{(\beta,\gamma)}   
    (t)  
\end{split}  \label{eq:result3}
\end{equation}
obtained by eliminating $P_{n-1}^{(\beta,\gamma)}(t)$ between the Jacobi recursion and differential relations (see equations (22.7.1) and (22.8.1) of \cite{abramowitz}). Hence equation (\ref{eq:PDM-psi-bis}) is proved.\par
% 
%--------------------------------------------------------------------------------------------------------
%
{}Finally, on combining equations (\ref{eq:result1})--(\ref{eq:result3}), we arrive at a recursion relation for the normalization coefficient
\begin{equation*}
  \frac{\na_{n+1,L}}{\na_{n,L}} = \left(\frac{(n+1) \left(n + \lambdaa + L + 1\right) \left(2n + \lambdaa
  + L + 3\right)}{\left(n + L + \frac{3}{2}\right) \left(n + \lambdaa + \frac{1}{2}\right) \left(2n +
  \lambdaa + L + 1\right)}\right)^{1/2}
\end{equation*}
whose solution is given by (\ref{eq:PDM-norm}). This completes the determination of the wavefunctions $\psia_{n,L}(r)$.\par
%
%=============================================================
%

\end{document}